\documentclass[twocolumn,aps,floatfix,showpacs,superscriptaddress]{revtex4}

\usepackage{amsmath,amssymb,eucal,graphicx}
\usepackage{color}
\usepackage{ulem}

\begin{document}

\title{Critical behavior of the exclusive queueing process}

\author{Chikashi Arita}
\affiliation{Institut de Physique Th\'{e}orique, CEA Saclay, 
91191 Gif-sur-Yvette, France}
\affiliation{ Theoretische Physik, Universit\"{a}t des Saarlandes, 
 66041 Saarbr\"ucken, Germany  }
\author{Andreas Schadschneider}
\affiliation{  Institut f\"{u}r Theoretische Physik,
 Universit\"{a}t zu K\"{o}ln, 50937 K\"{o}ln, Germany}


\begin{abstract}
The exclusive queueing process (EQP) is a generalization of
  the classical M/M/1 queue. It is equivalent to a totally asymmetric
  exclusion process (TASEP) of varying length. Here we consider two
  discrete-time versions of the EQP with parallel and
  backward-sequential update rules.  The phase diagram (with respect
  to the arrival probability $\alpha$ and the service probability
  $\beta$) is divided into two phases corresponding to divergence and
  convergence of the system length.  We investigate the behavior on
  the critical line separating these phases.  For both update rules,
  we find diffusive behavior for small service probability ($\beta <
  \beta_c$).  However, for $\beta>\beta_c$ it becomes sub-diffusive
  and nonuniversal: the critical exponents 
 characterizing   the divergence
  of the system length and the number of customers are found to depend
  on the update rule.  For the backward-update case, they also depend
  on the hopping parameter $p$, and remain finite when $p$ is large,
  indicating a first order transition. 
 \end{abstract}

\maketitle

\section{Introduction}

The M/M/1 queueing process describes the dynamics of a queue which is
specified by the arrival probability $\alpha$ and service probability
$\beta$ \cite{ref:Medhi,ref:Saaty}.
It has two phases separated by the {\it critical line} $\alpha=\beta$: for
 $\alpha>\beta$  
the length of the queue diverges whereas it converges
for  $\alpha<\beta$.
In the M/M/1 queueing process, the internal structure of the queue is
not considered, i.e. the queue 
has density 1 everywhere.

The exclusive queueing process (EQP) is a simple generalization of
this classical M/M/1 queueing process. It was introduced to impose
excluded-volume effect such that the internal structure of queues is
taken into account \cite{ref:A,ref:YTJN,ref:AY}.  Customers (i.e.
``particles'') move according to the rules of the totally asymmetric
simple exclusion process (TASEP), which is a paradigmatic model of
interacting many-particle systems far from equilibrium \cite{ref:D,ref:SCN}.  
The EQP can be regarded as a TASEP with 
variable system length.  Some TASEPs or related systems
with a variable length have been studied, especially in the context
of biological applications, e.g. to the growth of hyphae,
microtubules or bacterial flagellar filaments
\cite{ref:SEPR,ref:SE,ref:ES,ref:DMP,ref:JEK,ref:MRF,ref:SS}.

\begin{figure}
\includegraphics[width=8cm]{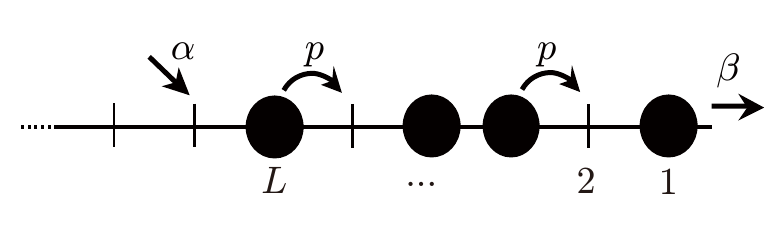}
\vspace{-0.9cm}
\caption{Dynamics of the exclusive queueing process (EQP).}
\label{fig:model}
\end{figure}

The EQP is defined on a semi-infinite one-dimensional lattice, where
the sites are labeled by natural numbers from right to left, see
fig.~\ref{fig:model}.  Each site $j$
 is either empty or occupied by a customer.
The system length $L$ is defined as the position of the leftmost
occupied site.  It is in general different from the number $N$ of
customers which have unit length in contrast to the M/M/1 queue
where one has always $L=N$.
Due to the exclusion principle, customers move forward with
probability $p$ in each time step only if the preceding site is
unoccupied.  A new customer enters the system at the end of the queue,
i.e. at the site $L+1$ next to the leftmost occupied site, with
probability $\alpha$.  The customer at the rightmost site $j=1$ gets
service with probability $\beta$ and is removed from the system.

We need to specify an update rule to fully define the dynamics of the
EQP. Here we consider parallel and backward-sequential update rules.
In the EQP with parallel update ({\it parallel EQP}), all sites
are updated simultaneously.  In the EQP with backward-sequential
update ({\it backward EQP}), first a customer arrives with probability
$\alpha$, and the customer at the right end ($j=1$) is extracted with
probability $\beta$ (if it exists).  Then starting from the rightmost
customer and going sequentially to the left up to the leftmost
customer, the system is updated according to the rules of the TASEP
(see \cite{ref:AS3} for more details)
One can also consider the EQP with continuous time \cite{ref:A}.
Relations among the two discrete-time EQPs, the continuous-time EQP,
and some special cases have been studied in \cite{ref:AS3}.

Exact stationary states for the continuous-time and parallel EQPs have
been found \cite{ref:A,ref:AY}.  However, obtaining an exact ``dynamical
state'' (time-dependent solution) was not possible so far except for
deterministic hopping $p=1$ \cite{ref:AS1,ref:AS2}.  Thus we rely on
Monte Carlo simulations to investigate time-dependent properties of the
EQPs.  In this work, we focus on critical properties,
i.e. the behavior of the system length $L$ and 
the number $N$ of customers on
the critical line of the EQP.
Getting reliable results then requires
averaging over a large number of samples and long times.

\section{Critical line}

The critical line that separates the convergent and divergent phases
in the usual M/M/1 queueing process is simply $\alpha=\beta$.  
In the EQP case, it is modified depending on the update rule
\cite{ref:AS1,ref:AS3}: for the parallel EQP,
\begin{equation}
\alpha_c  = 
 \begin{cases}
 \frac{\beta(p-\beta)}{p-\beta^2} & \qquad\text{for\ }\beta\le\beta_c,\\ 
 \frac{1-\sqrt{1-p}}{2}           & \qquad\text{for\ }\beta>\beta_c,
 \end{cases}
\label{eq:alphac-para}
\end{equation} 
and for the backward EQP,
\begin{equation}
\alpha_c= \begin{cases}   
  \frac{\beta(p-\beta)}{p(1-\beta)}  & \qquad\text{for\ }\beta\le\beta_c,  \\
  \frac{(1-\sqrt{1-p})^2 }{p}        & \qquad\text{for\ }\beta>\beta_c,
\end{cases}
\label{eq:alphac-back}
\end{equation} 
where  
\begin{equation}
\beta_c=1-\sqrt{1-p} 
\end{equation}
is independent of the update rule.
The form for $\alpha_c$ corresponds to the outflow of customers, 
and thus the time-dependent behavior of the average number of customers
is well expressed as \cite{ref:AS1} 
\begin{equation}\label{eq:Nt-sim}
\langle N_t\rangle\sim (\alpha-\alpha_c) t 
\end{equation}
which is the asymptotic form in the divergent phase
($\alpha>\alpha_c$).  Equation (\ref{eq:Nt-sim}) is true only for
$t\lesssim ( \alpha_c -\alpha )/\langle N_0\rangle$ in the convergent
phase ($\alpha<\alpha_c$) where $\langle N_t\rangle $ converges to a
stationary value.  Similarly, the length $\langle L_t \rangle$ of the
system converges to a stationary value ($\alpha>\alpha_c$) or 
 diverges linearly in $t$ ($\alpha<\alpha_c$).

\begin{figure}
 \includegraphics[width=7.5cm]{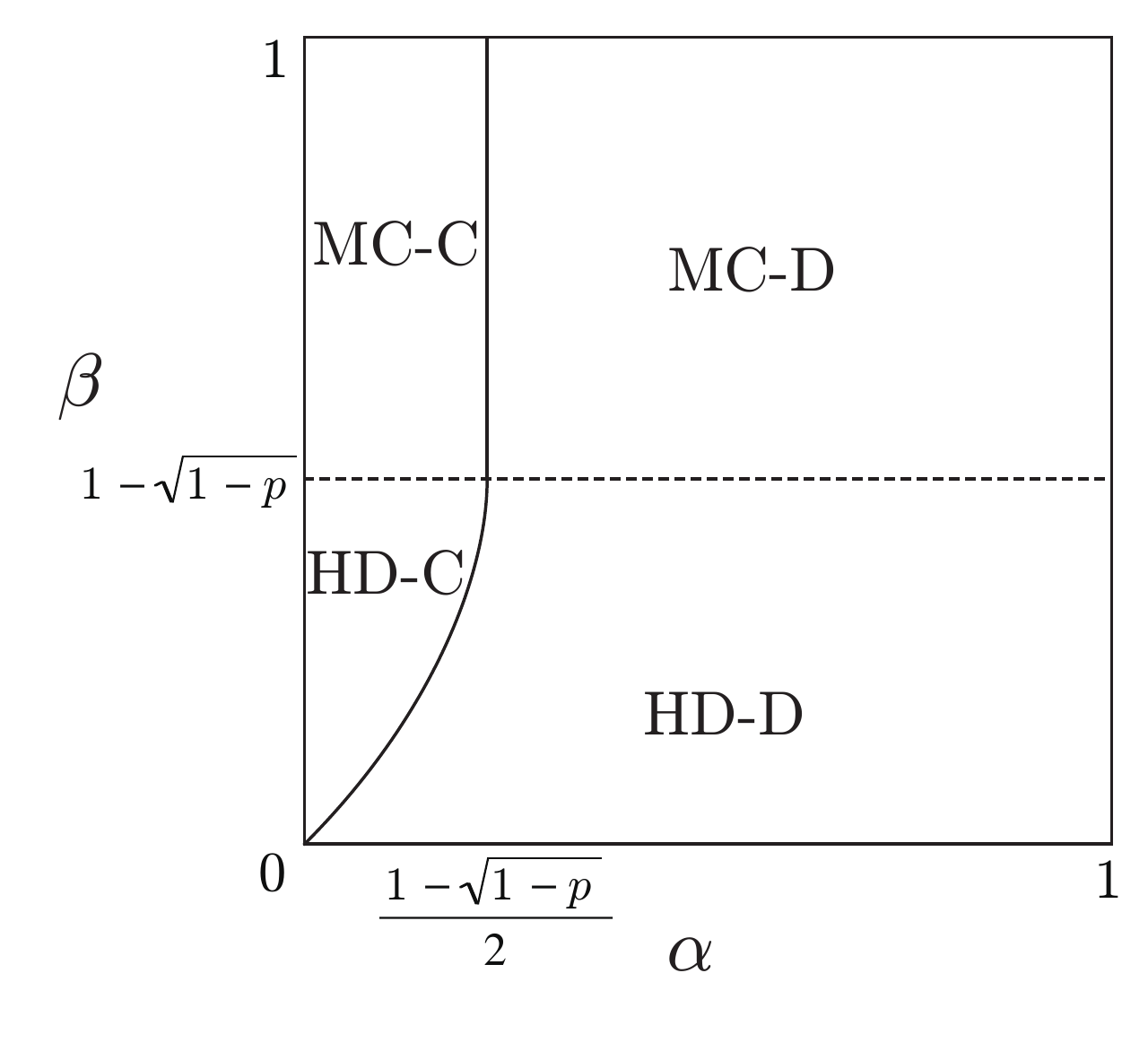}\qquad
\vspace{-0.5cm}
\caption{Phase diagram of the  parallel EQP.
The phase diagram of the backward EQP is qualitatively the same, 
with $\frac{1-\sqrt{1-p}}{2}$ replaced by $\frac{(1-\sqrt{1-p})^2}{p}$.
The phases with diverging length are indicated by the extension ``-D''
and convergent phases by ``-C''.  The maximal current (MC) and
high-density (HD) phases are distinguished by the form of $\alpha_c$
in (\ref{eq:alphac-para}) and (\ref{eq:alphac-back}).  }
\label{fig:phasediagram}
\end{figure}

The results (\ref{eq:alphac-para}) and (\ref{eq:alphac-back}) suggest
a division of the phase diagram into four phases
(fig.~\ref{fig:phasediagram}) by distinguishing between maximal
current and high density phases \cite{ref:AS1}.  The divergent phase
is further divided into up to five subphases 
according to the shape of the
density profile.  The pair of coefficients (or growth velocities)
$\bigl(\frac{\langle N_t\rangle}{t}$, $\frac{\langle
  L_t\rangle}{t}\bigr)$ has a different expression in each subphase
\cite{ref:AS3}.

\begin{figure}
\vspace{-2mm}

\includegraphics[width=8cm]{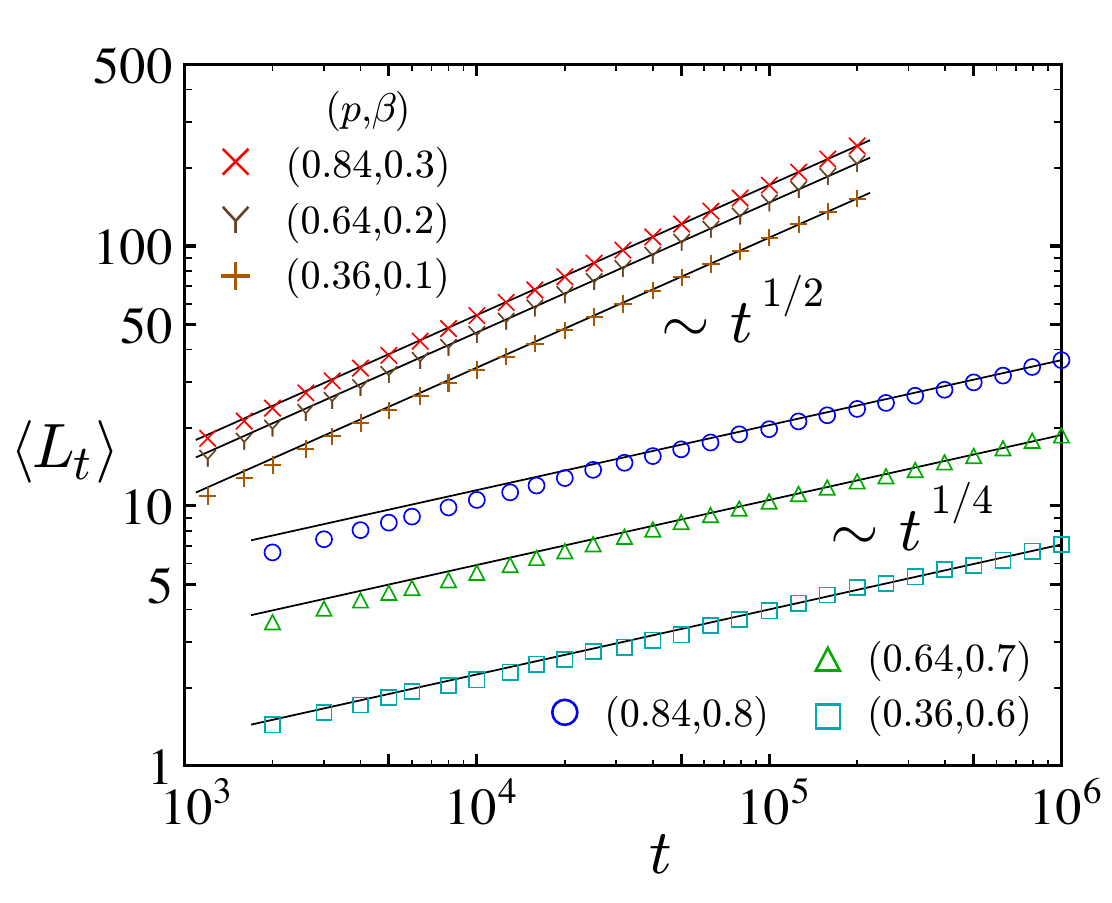}
\vspace{-2mm}

\includegraphics[width=8cm]{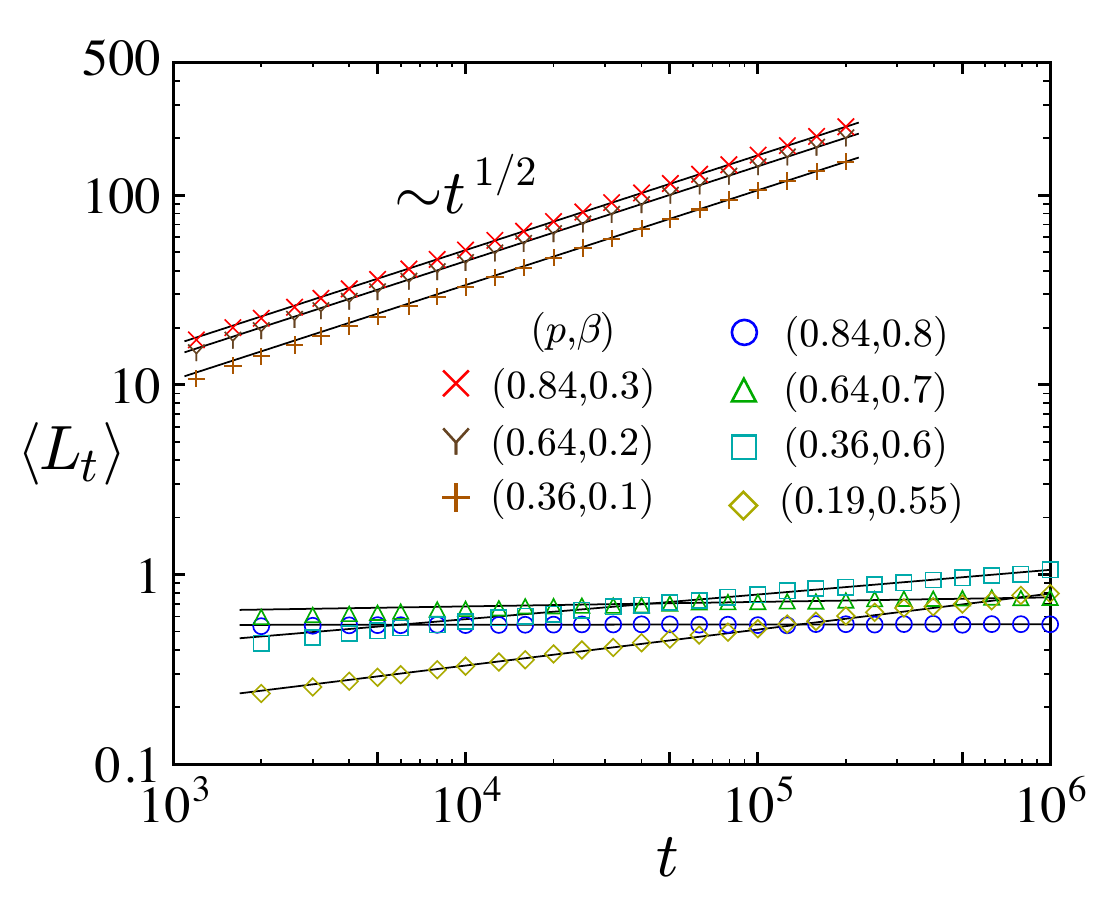}
\vspace{-5mm}

\caption{
  Log-log plots
  of the average system length $\langle L_t \rangle $ for the parallel
  (top) and backward (bottom) EQPs.  Simulations were
    averaged over $10^5$ and $10^6$ samples for the curved part
    ($\times$, Y, $+$) and the straight part
    ($\bigcirc,\triangle,\square,\Diamond$) of the critical line,
    respectively.  The data are in good agreement with the power law 
    behavior (\ref{eq:X=tgamma}).   To obtain the fitting lines,
    we fixed $\gamma_L = \frac{1}{2}$ or $ \frac{1}{4}$ except for the
    backward case on the straight part,
        where the exponents have been estimated from
        eq. (\ref{eq:log/log}).
  }
\label{fig:loglog}
\end{figure}

In our previous work \cite{ref:AS3}, we have observed the behavior
\begin{equation}\label{eq:X=tgamma}
\langle X_t \rangle = O\left( t^{\gamma_X}\right)   
\end{equation}
of   the system  length ($X=L$) and 
 the number of customers  ($X=N$)  
  just on the critical line
with non-trivial {\it growth exponents} $\gamma_X$. 
Since clearly $\gamma_X=0$ in the convergent phase and $\gamma_X=1$ in
the divergent phase, it is natural to expect $0\le \gamma_X\le 1$ on
the critical line. This is indeed confirmed by the simulations,
see figs.~\ref{fig:loglog},~\ref{fig:gamma}.
(In fig. ~\ref{fig:gamma}, we rescale the service probability
$\beta$ such that $\tilde\beta=0,0.5$ and 1 
correspond to $\beta=0,\beta_c$ and 1, respectively.)

Here, we will examine the exponents systematically, and find that they
exhibit different behavior depending on parts of the critical line
(curved part $\beta<\beta_c$, i.e. the phase boundary between the
high-density subphases, or straight part $\beta>\beta_c$, i.e.  the
phase boundary between the maximal-current subphases).  To obtain
reliable results, we take averages over a large number ($10^5$, $10^6$
or more) of simulation samples with up to $10^6$ time steps.  In
particular for the backward case, fluctuations are strong and a large
number of samples is required to determine the exponents accurately.
As initial condition ($t=0$), simulations are started from an empty
lattice where no customers are present in the system.

\section{On the curved part}

For the case of deterministic hopping case $p=1$ rigorous results
exist \cite{ref:YTJN,ref:AS1,ref:AS2}.  In this case,
$\beta_c=1$ and the MC-C and MC-D phases vanish from the phase diagram.  
For the parallel EQP one has $L\neq N$ due to the exclusion principle
and a time-dependent solution can be obtained in matrix
product form \cite{ref:AS2}.  On the other hand, the backward EQP with
$p=1$ reduces to the usual M/M/1 queue with $L=N$.  On the critical
line, i.e.
\begin{eqnarray}\label{eq:alpha=}
 \alpha =  \begin{cases}
\frac{\beta}{1+\beta} &  \qquad ({\rm parallel}) ,\\
\beta                 &  \qquad ({\rm backward}) ,
\end{cases}
\end{eqnarray}
the system length and the number of customers exhibit diffusive behavior:
\begin{eqnarray}
\label{eq:p1-crit}
   \langle L_t\rangle  = C_L \sqrt{t}  +o(\sqrt{t}), \qquad 
   \langle N_t\rangle  = C_N \sqrt{t}  +o(\sqrt{t}) ,
\end{eqnarray}
where the coefficients depend on the update rules,
\begin{eqnarray}\label{eq:C_X}
  C_L =  
 \left\{
 \begin{array}{ll}
  2\sqrt{ \frac{\beta(1-\alpha)}{ \pi } }  & \ ({\rm parallel}), \\ 
  2\sqrt{ \frac{\beta(1-\beta)}{\pi} }     & \ ({\rm backward}), 
 \end{array}\right.  
 \qquad  C_N =  \rho\, C_L, 
\end{eqnarray}
with the average density  
\begin{eqnarray}\label{eq:rho:p=1}
 \rho =  \begin{cases}
\frac{1}{1+\beta}  &  \qquad ({\rm parallel)},\\
1                  &  \qquad ({\rm backward}).
\end{cases}
\end{eqnarray}
Note that since $\alpha$ and $\beta$ are related by equation
(\ref{eq:alpha=}), one can express the coefficients in various ways.

We denote the probability that site $j$ is occupied by a customer at
time $t$ by $\rho_{jt}$.  For the deterministic hopping case $p=1$,
this density profile can be expressed by the complementary error
function ${\rm erfc}(x) =\int^{\infty}_{x}e^{-y^2}dy$ as
\begin{eqnarray}\label{eq:rhojt=}
\rho_{x\sqrt{t},t}  \to  \rho\ 
{\rm erfc} \left( \frac{\sqrt{\pi} }{ C_L} \right)\qquad
(t\to\infty).
\end{eqnarray}

We now turn to the behavior on the curved part for general $p$.  The
exponents $\gamma_X$ ($X=L,N$) are estimated from the
simulation data by
\begin{equation}\label{eq:log/log}
 \ln  \frac{\langle X_{t}\rangle}{\langle X_{t/10}\rangle }
\Big/{\ln 10}
\end{equation}
which approaches the true exponent $\gamma_X$ for $t\to\infty$.

\begin{figure}
\begin{center}
\vspace{-12mm}

 \includegraphics[width=8cm]{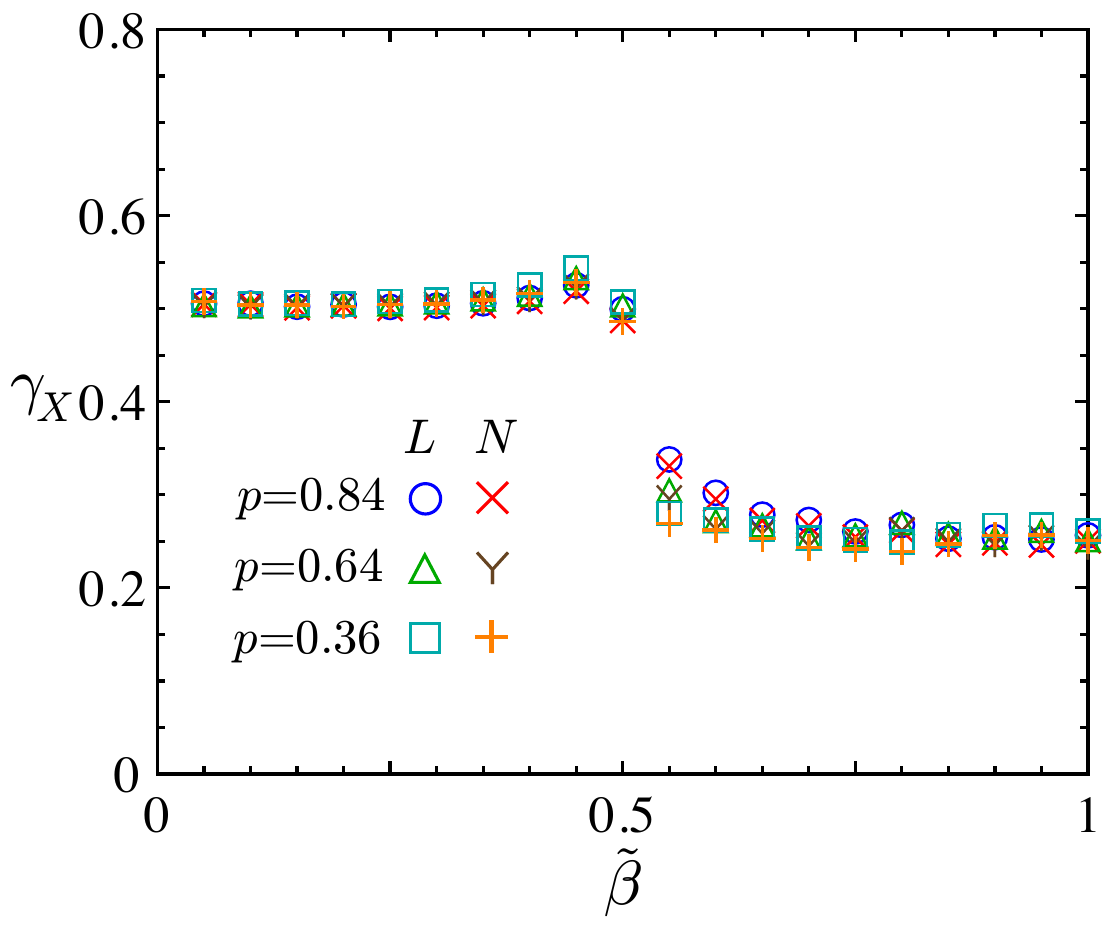}
\vspace{-12mm}

 \includegraphics[width=8cm]{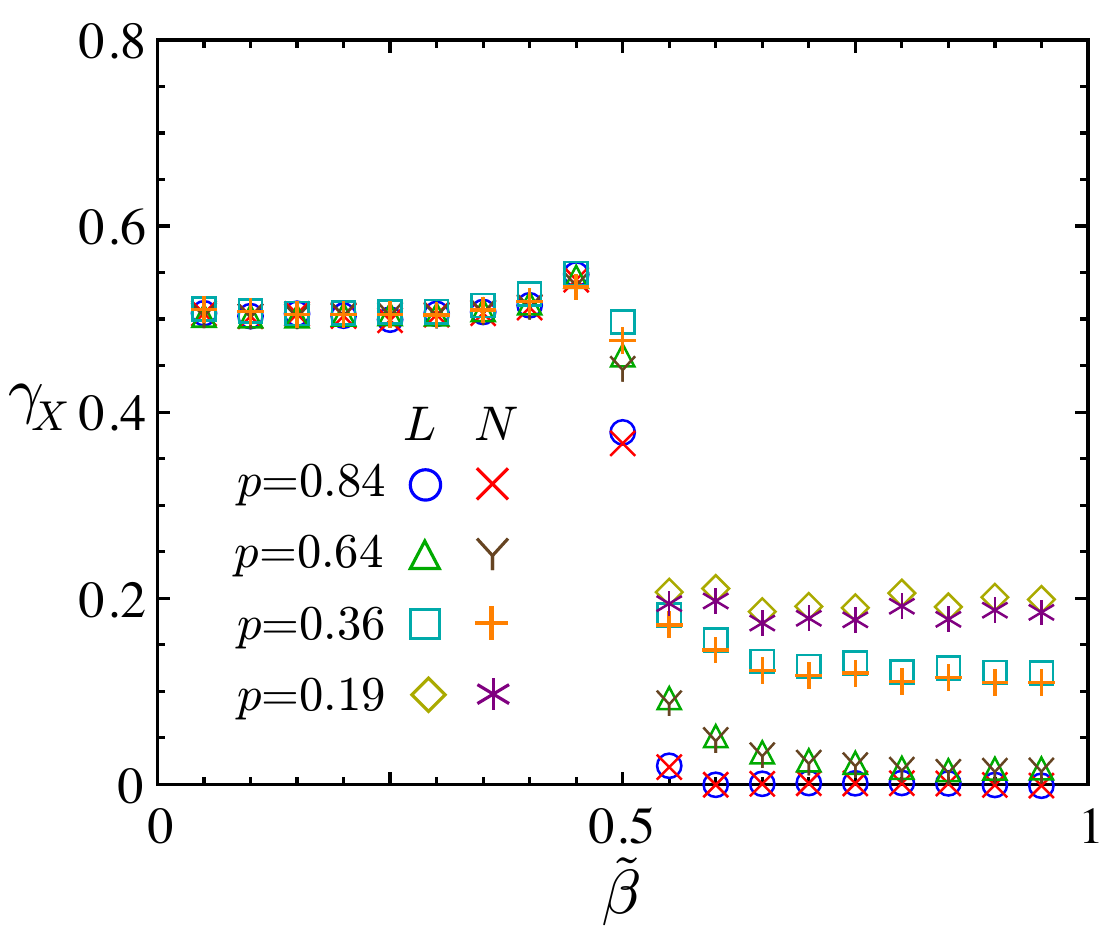}
\end{center}
\vspace{-7mm}

\caption{Growth exponents $\gamma_X$ for the parallel (top) and backward
  (bottom) EQPs based on estimates from eq.~(\ref{eq:log/log}).
  The parameter $\beta$ has been rescaled as $\tilde{\beta} =
  \frac{\beta}{2 \beta_c}$ (for $\beta\le\beta_c),
  \frac{\beta+1-2\beta_c}{2(1-\beta_c)}$ (for $\beta>\beta_c)$ for
  better visibility.  The separation of the straight and the curved
  part of the critical line is then located at $\tilde{\beta}
  =\frac{1}{2}$.
  The hopping parameter is chosen as $p=0.84,0.64,0.36$ (additionally
  $p=0.19$ for $\tilde\beta>1/2$ in the backward case).
  Simulations up to $t\le 2\times 10^5$ and $t\le10^6$ for $\tilde
  \beta \le 0.5$ and $\tilde \beta > 0.5$, respectively, have been
  used with averaging over $10^5$ samples.  On the straight part for
  the backward case, averaging was done over $(2\tilde\beta-1)\times
  10^6$ (for $\tilde\beta< 0.75) $ or $10^6$ (for $\tilde\beta\ge
  0.75) $ samples due to the strong fluctuations.  
  Clearly the exponents are different for $\tilde \beta < 0.5$
  and $\tilde \beta >0.5$.}
\label{fig:gamma}
\end{figure}

The results in fig.~\ref{fig:gamma} strongly indicate that
on the curved part ($\beta < \beta_c$, i.e. $\tilde\beta<0.5$)
the exponents are given by
\begin{equation}
\gamma_X=\frac{1}{2} \,,
\end{equation}
i.e. diffusive behavior as in the deterministic case $p=1$.

\begin{figure}
\begin{center}
 \includegraphics[width=8cm]{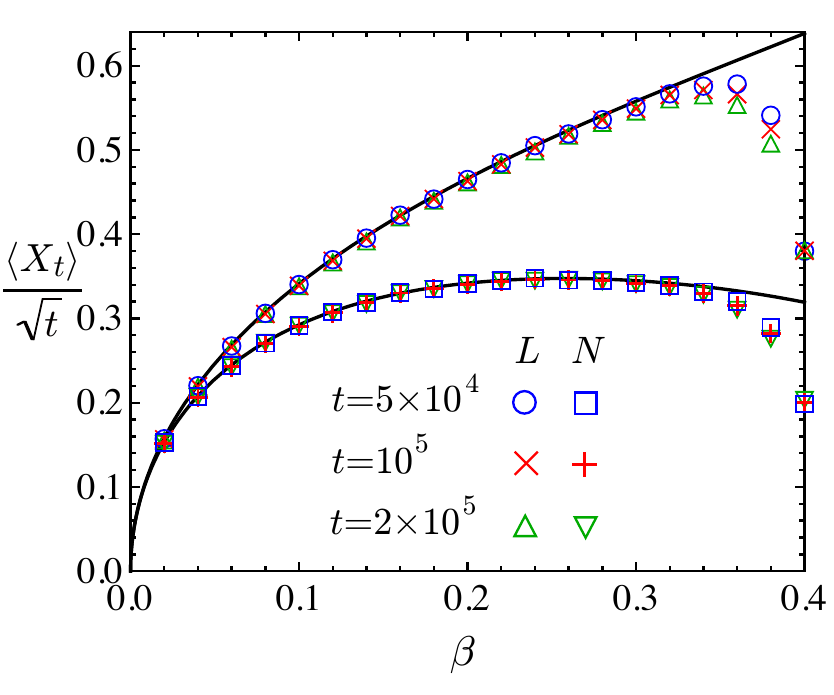}
  \vspace{-2mm}
 \includegraphics[width=8cm]{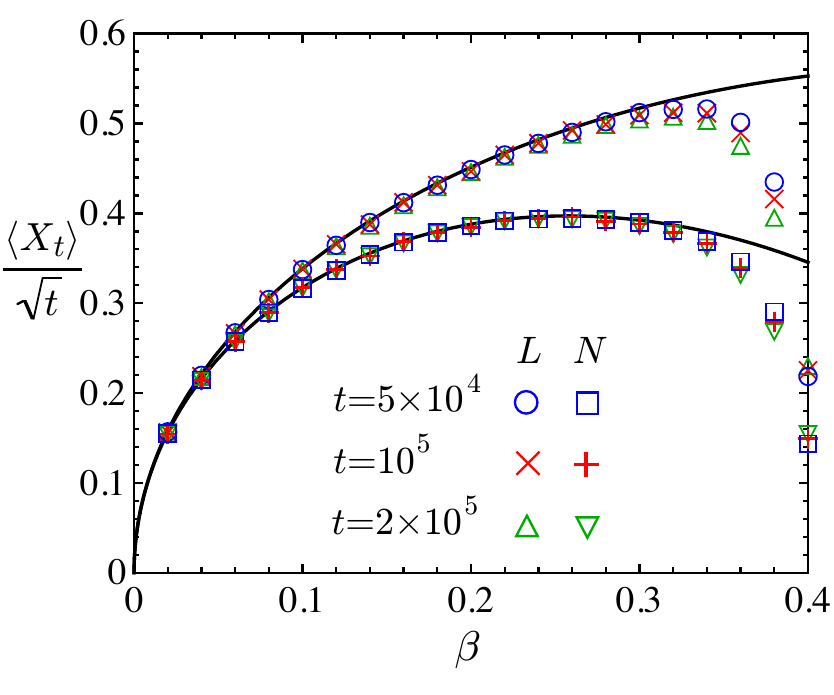}
\end{center}
\vspace{-0.5cm}
\caption{
  The coefficients $C_X=\frac{\langle X_t\rangle}{\sqrt{t}}$ ($X=L,N$)
  on the curved part of the critical line for the parallel (top) and
  backward (bottom) EQPs.  The hopping probability is chosen as
  $p=0.64$, where $\beta_c=0.4$.  The markers are plotted by averaging
  $10^5$ simulation samples.  We observe that they agree with $C_L$
  and $C_N$ defined by eqs.~(\ref{eq:C_X}) and (\ref{eq:rho})
  (full lines) when $p$ is small.}
\label{fig:Coeff}
\end{figure}

\begin{figure}
\begin{center}
 \includegraphics[width=8cm]{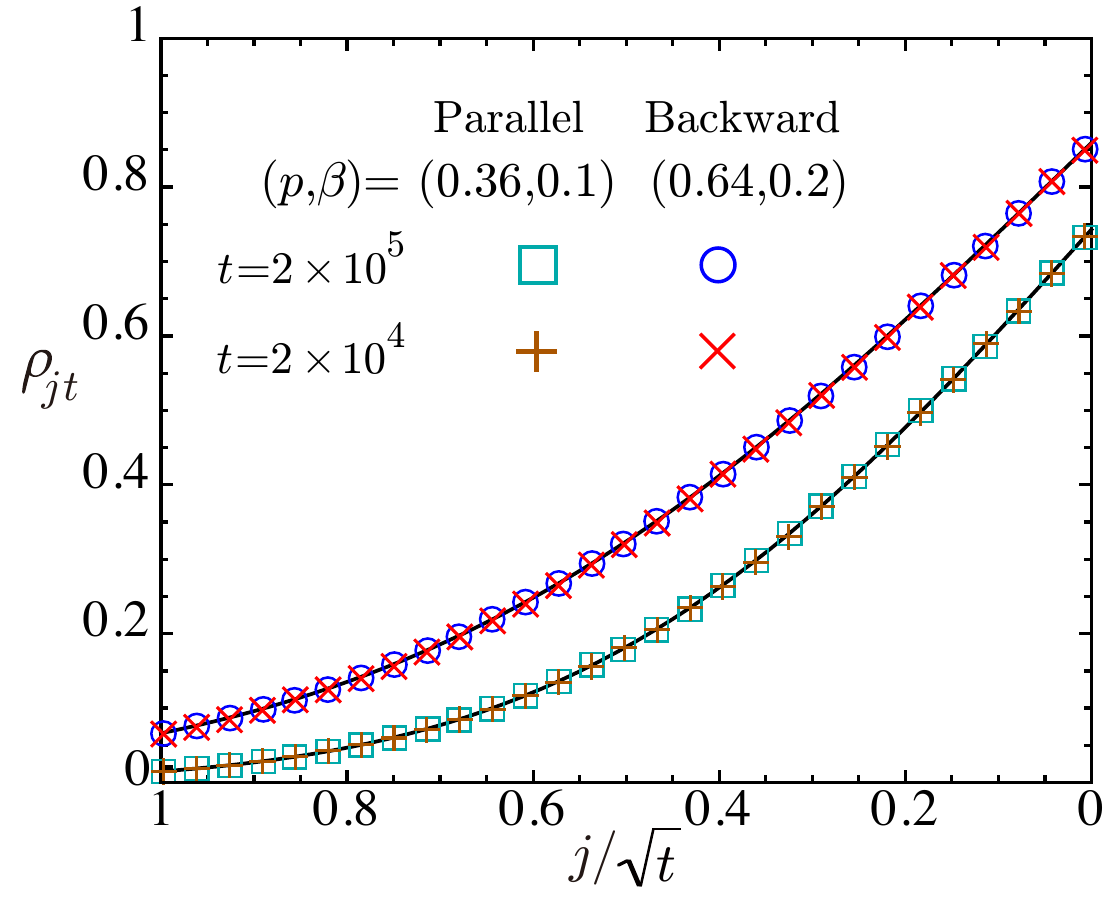}
\end{center}
\caption{The rescaled density profiles on the curved part 
  of the critical line for the parallel and backward
  EQPs.
  The plots were  obtained by averaging $10^5$ simulation samples.
  They agree with
  the lines corresponding to (\ref{eq:rhojt=}) with
  (\ref{eq:C_X}) and  (\ref{eq:rho}).
\label{fig:rho}
}
\end{figure}

Next we estimate the coefficients $C_X$. Interestingly, simulation 
data for $\frac{\langle L_t\rangle}{\sqrt t}$ are in good agreement 
with the exact result (\ref{eq:C_X}) for $C_L$ in the deterministic
case $p=1$ except near $\beta_c$ (fig.~\ref{fig:Coeff}).
In a similar way, the form (\ref{eq:C_X}) for $\frac{\langle
  N_t\rangle}{\sqrt t}$ with a modification of the mean density as
\begin{eqnarray}\label{eq:rho}
 \rho =  \begin{cases}
\frac{p-\beta}{p-\beta^2}  &  \qquad ({\rm parallel}),\\ 
\frac{p-\beta}{p(1-\beta)} &  \qquad ({\rm backward}), 
\end{cases}
\end{eqnarray}
fits simulation results well (fig.~\ref{fig:Coeff}).
Furthermore the form (\ref{eq:rhojt=}) gives a good expression for
rescaled density profiles (fig.~\ref{fig:rho}).

These facts imply that the EQPs on the critical line are described by
noninteracting random walkers hopping rightward or leftward
 with the same probability $ \frac{ \pi }{4}C^2_L$. 
This is exactly true for the
deterministic case $p=1$ \cite{ref:AS2}.

\begin{figure}
\begin{center}
 \includegraphics[width=8cm]{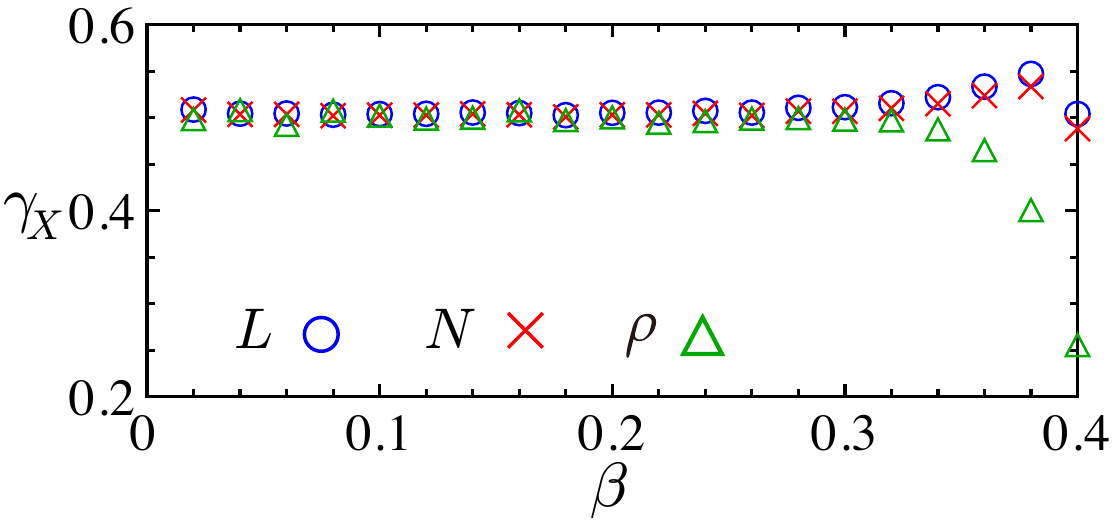}
 \includegraphics[width=8cm]{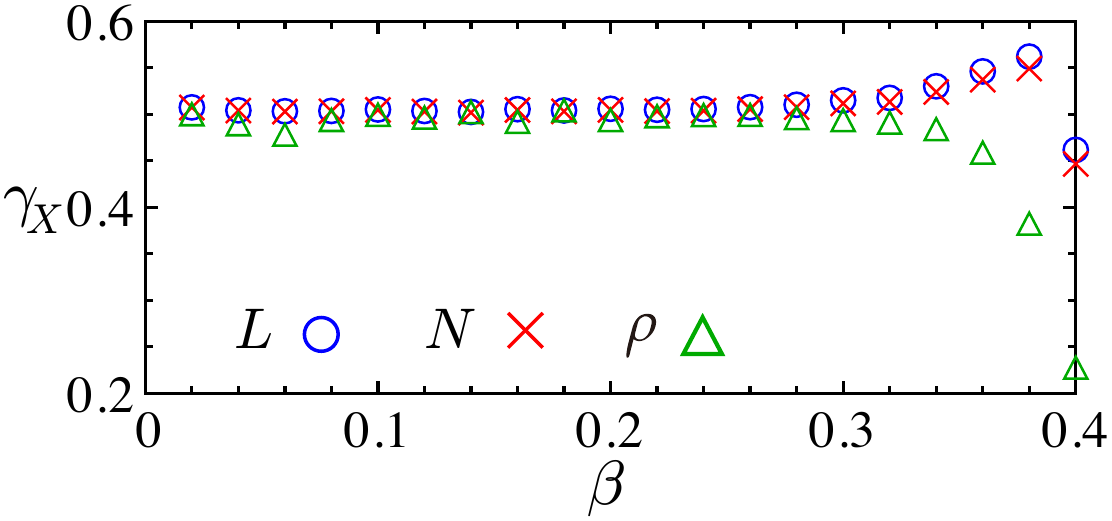}
\end{center}
\vspace{-0.5cm}
\caption{The exponents $\gamma_X$ ($X=L,N,\rho$) for the parallel 
  (top) and backward (bottom) EQPs
  on the curved part of the critical line with $p=0.64$.
   We applied eqs.~(\ref{eq:log/log}) and
  (\ref{eq:log/log-rho}) to $10^5$  
   samples up to $t=2\times 10^5$ for the estimations.}
\label{fig:gamma-cv}
\end{figure}

In fig.~\ref{fig:Coeff}, we observe that the finite-time effects
become larger as $ \beta\nearrow\beta_c$, i.e. $\langle
X_t\rangle/\sqrt{t}$ approaches $C_X$ more slowly.  This effect can
also be observed on the level of the exponents, see
fig.~\ref{fig:gamma-cv}.  We observe that the exponents $\gamma_L$ and
$\gamma_N$ are shifted upward near $\beta_c$.  Fig.~\ref{fig:gamma-cv}
also shows the exponent $\gamma_\rho$ of the mean density which is
defined by
\begin{equation}\label{eq:gamma-rho}
  \frac{\langle N_t \rangle}{\langle L_t \rangle} -  \rho
   =  O(t^{-\gamma_\rho}) 
\end{equation}
with the limit density (\ref{eq:rho}). It can be estimated by using a
formula, which is similar to (\ref{eq:log/log}),
\begin{equation}\label{eq:log/log-rho}
\ln\frac{\langle N_{10t} \rangle/\langle L_{10t} \rangle- \rho}
 {\langle N_t \rangle/\langle L_t \rangle-  \rho}
\Big/{\ln 10}\,.
\end{equation}
This exponent is expected to be identical to the two growth exponents
$\gamma_L$ and $\gamma_N$, but the finite-time effect shifts it
downward near $\beta_c$.

\section{On the straight line}

We first consider the parallel case.  The top graph of
fig.~\ref{fig:gamma} shows the exponents $\gamma_L$ and $\gamma_N$ for
the parallel update.  It indicates subdiffusive behavior
  for $\beta>\beta_c$,
i.e. $\tilde \beta >0.5$.  We expect
$\gamma_L=\gamma_N$ so that the total density $\rho_{\rm tot}=\langle
N_t \rangle/\langle L_t\rangle$ reaches the finite value
\begin{equation}
  \rho =
   \frac{1}{2}                \qquad ({\rm parallel}), 
\label{eq:rho=}
\end{equation} 
which corresponds to the  density of the 
maximal current  \cite{ref:RSSS}. 
Although we observe a tendency that $\gamma_L$ is slightly larger
than $ \gamma_N$, this can be considered to 
be a systematic finite-size effect.

\begin{figure}
\begin{center}
 \includegraphics[width=8cm]{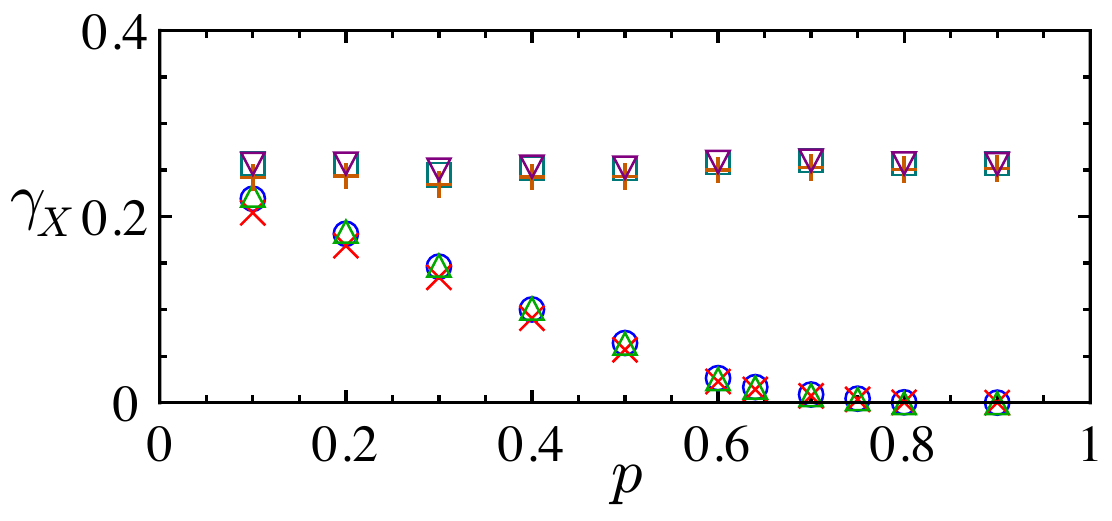}
 \includegraphics[width=8cm]{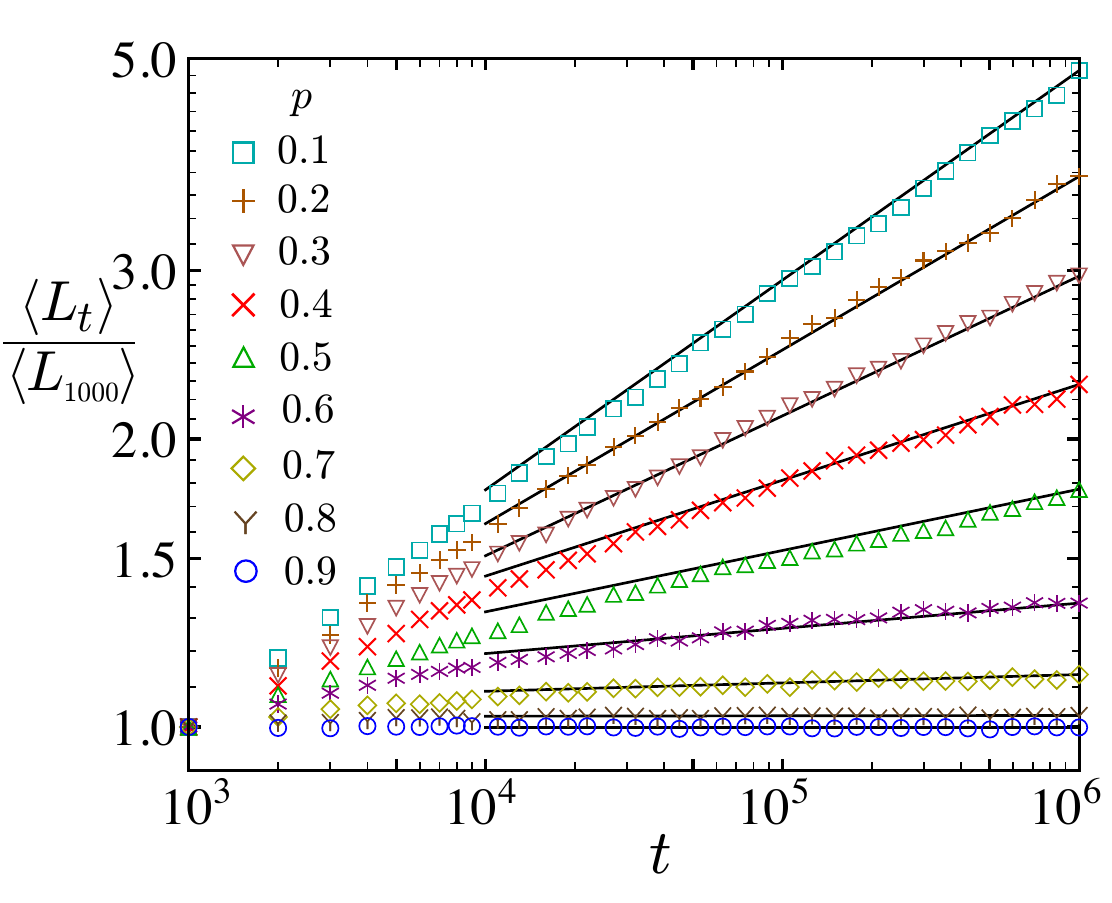}
\end{center}
\vspace{-0.5cm}
\caption{ 
(Top)  the exponents $\gamma_X$ on the straight part of the critical lines
  of the parallel ($X=L(\square),N(+),\rho(\bigtriangledown)$) and
  backward ($X=L(\bigcirc),N(\times),\rho(\bigtriangleup)$) EQPs.
   (Bottom) log-log plots of the average system length
   (normalized by  $\langle L_{10^3}\rangle $ 
  for better visibility)
  on the straight part of the critical line of the 
    backward EQP. 
We have set  $\beta=(\beta_c+4)/5$
with various values of $p$.
  We applied eqs.~(\ref{eq:log/log}) and
  (\ref{eq:log/log-rho}) to $10^6$ (parallel) and $5\times 10^6$
  (backward) samples up to $t=10^6$ for the estimations.
with    $\beta=(\beta_c+4)/5$.   }
\label{fig:gamma-straight}
\end{figure}

The results shown in the top graph of fig.~\ref{fig:gamma}
are compatible with universal behavior with the exponents
\begin{equation}
\gamma_L=\gamma_N=\frac{1}{4}  
\end{equation}
for the parallel case.  This is further supported by
fig.~\ref{fig:gamma-straight} where $\gamma_X$ are
shown with  $\beta=(4+\beta_c)/5$
and various values of the hopping probability $p$.
Furthermore the exponent for the mean density  (\ref{eq:gamma-rho}) 
is expected to be identical to those for $L$ and $N$:
\begin{equation}
 \gamma_L =   \gamma_N = \gamma_\rho \,.
\label{eq:gL=gN=grho}
\end{equation} 
For the curved part, we have seen that the behavior
  is well described by a  symmetric random walk
  model.  We expect that the exponent $\frac{1}{4}$ will also
 be understood by mapping to a simple model, 
  which we leave as an open problem.   

Let us turn to the straight part in the backward case
($\tilde{\beta}>0.5$ in the bottom graph of fig.~\ref{fig:gamma}).
For $\gamma_\rho$, we use  eq.~(\ref{eq:log/log-rho})  with 
\begin{align}\label{eq:lim-rho-back}
\rho =\frac{1-\sqrt{1-p}}{p} \qquad ({\rm backward}) .
\end{align}
Surprisingly the critical behavior turns out to be rather different
from that for the parallel case.  Although the exponents $\gamma_X$
($X=L,N,\rho$) are identical to each other
 (see eq.~(\ref{eq:gL=gN=grho}))
for each choice of the parameters, they depend on $p$ (but are
independent of $\beta$) and thus the behavior on the straight part is
nonuniversal.

 When $p$ is small,
$L$ and $N$ seem to continue to grow with a power law,
see the bottom graph of fig.~\ref{fig:gamma-straight}. 
From the top graph of fig.~\ref{fig:gamma-straight},
we find  $\gamma_X\to\frac{1}{4}$ as $p\to 0$, which matches the exponent
for the parallel case.
When $p$ is large, we cannot find conclusive evidence for a
divergence of $L$ and $N$,
see again the bottom graph of fig.~\ref{fig:gamma-straight}.
Note that in the limit $p\to 1$ (usual M/M/1  case) the straight line part
in the phase diagram shrinks to a point $\alpha=\beta=1$.
There we can easily show that only the empty chain is realized, i.e.
$\langle L_t \rangle =0$, which matches the results for large $p$.
This property is different from the parallel case, i.e. the straight
line shrinks to  just the point  $(\alpha,\beta)=(0.5,1)$ 
in the limit $p\to 1$, where $L$ grows infinitely \cite{ref:AS2}.

Assuming that $\gamma_X$ takes non-zero values when 
$p$ is small, and  $\gamma_X=0$ when $p$ is large,
there exists a point $p_c\approx 0.7$ such that 
\begin{equation}
\gamma_X \begin{cases}
>0 & \qquad \text{for }p<p_c,\\
=0 & \qquad \text{for }p\ge p_c.
\end{cases}
\label{eq:p_c}
\end{equation}
The exponent  $\gamma_L=0$ ($p\ge p_c$)
indicates that the transition from the
convergent to the divergent phase is of first
order, which can be seen more clearly by introducing
\begin{equation}
m=\lim_{t\to\infty}   \langle L_t\rangle ^{-1}
\end{equation}
as an order parameter.  In the divergent phase this parameter vanishes
($m=0$) whereas it stays finite in the convergent phase ($m>0$).  On
the critical line with $\beta>\beta_c$ and $p\ge p_c$, it takes
nonzero values so that $m$ changes discontinuously in passing through
the critical line.
 Although eq.~(\ref{eq:lim-rho-back}) can be considered as 
the limit of the mean density for $p<p_c$, 
this is no longer true for $p \ge p_c$ where
$\lim_{t\to\infty}\frac{\langle N_t \rangle}{\langle L_t \rangle} 
 \neq \rho$.

Other scenarios are possible.  For example, $L$ and $N$ could converge
even for small $p$, but with an extremely long relaxation time.
Another possibility is that $L$ and $N$ always diverge with extremely
small but nonzero exponents $\gamma_X$ or more slowly than a power
law, e.g.  $\langle X_t\rangle = O( \ln t) $ which has been found in a
reverse-biased exclusion process with varying length \cite{ref:SS}.
However, we could not confirm these scenarios, and eq.~(\ref{eq:p_c})
is the most reasonable interpretation of our simulation 
results ($t\lesssim 10^6$).

\section{At the multicritical point} 

\begin{figure}
\includegraphics[width=8cm]{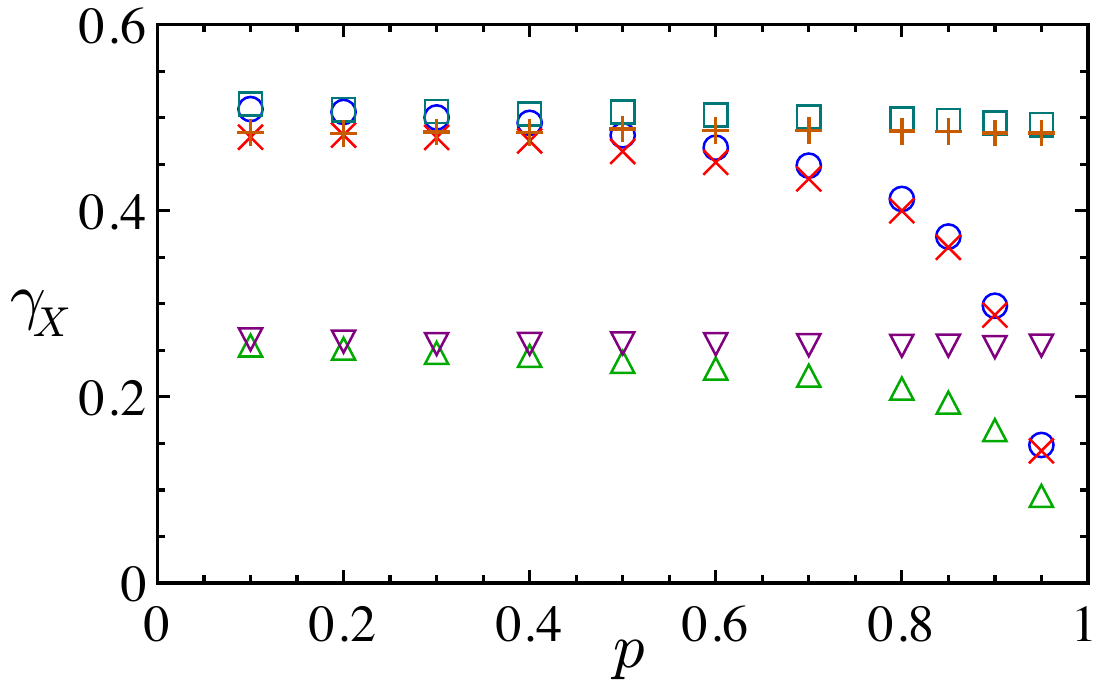}
\caption{The exponents $\gamma_X$   at the multicritical point 
  $(\alpha,\beta) = (\alpha_c,\beta_c)$ of the parallel
  ($X=L(\square),N(+),\rho(\bigtriangledown)$) and backward
  ($X=L(\bigcirc),N(\times),\rho(\bigtriangleup)$) EQPs with
  $\beta= \beta_c $.  We applied eqs.~(\ref{eq:log/log}) and
  (\ref{eq:log/log-rho}) to $10^5$ simulation samples up to $t=2\times
  10^5$ for the estimations.  }
\label{fig:gamma-point}
\end{figure}

We lastly examine the behavior at the multicritical point
$\beta=\beta_c$  where the straight and curved parts of the
critical line meet, see fig.~\ref{fig:gamma-point}.  For the parallel
EQP, we expect diffusive behavior $\gamma_L=\gamma_N=\frac{1}{2}$ as
found on the curved part.  However, the exponent for $\rho$ is not
identical to them, i.e.  $ \gamma_\rho=\frac{1}{4} $ as on the
straight part, so that (\ref{eq:C_X}) and (\ref{eq:rhojt=})
are not expected to be valid at the multicritical point.

In the backward case, the exponents again depend on $p$.
For small values of $p$, the exponents are almost the same as in the parallel case, whereas they become smaller as $p\to 1$.  
However, the dependences of $\gamma_\rho$ on $p$
at the multicritical point and on 
the straight part are different,
compare the plot markers $\triangle$ of 
 figs.~\ref{fig:gamma-straight}, \ref{fig:gamma-point}.

\section{Summary}

We have investigated the EQP, which is characterized by three
parameters (arrival probability $\alpha$, service probability $\beta$
and hopping probability $p$), with parallel and backward-sequential
update rules. In the $\alpha$-$\beta$ plane, phases of divergent and
convergent system length $L$ and customer number $N$ are separated by
a critical line which consists of a curved part for $\beta<\beta_c$
and a straight-line part for $\beta>\beta_c$.  Based on Monte Carlo
simulations, we have shown that on this critical line the growth
exponents $\gamma_X$ ($X=L,N$) are smaller than 1, the value in the
divergent phase \cite{ref:AS1}.  We introduced the exponent
$\gamma_\rho$ for the mean density, and we find generically
 $\gamma_L=\gamma_N=\gamma_\rho$.

More precisely, we find diffusive behavior $\gamma_X=\frac{1}{2}$
($X=L,N,\rho$) on the curved part ($\beta<\beta_c$) of the critical
line, which is independent of the update rule.  Based on exact results
in limiting cases, we also conjectured the coefficients (\ref{eq:C_X})
and the asymptotic form (\ref{eq:rhojt=}) of the rescaled density
profile, which agree well with the simulation results.

On the straight part ($\beta>\beta_c$) of the critical line, the
situation is not so simple. First of all, the behavior clearly depends
on the update rule.  For the parallel case, the exponents are found to
be in reasonable agreement with $\gamma_X=\frac{1}{4}$.  For the
backward case, however, the exponents depend on the hopping parameter
$p$.  The simulation results even indicate the existence of a point
$p_c$ such that $0<\gamma_X<\frac{1}{4}$ for $p<p_c$ whereas
$\gamma_X=0$ for $p\ge p_c$.  
This means that in this case the order of the transition on the
straight part changes from second order for small $p$ to first
order for large $p$.

 At the multicritical point $\beta=\beta_c$, 
we also found the nonuniversality and 
 $\gamma_L=\gamma_N\neq\gamma_\rho$.
For the parallel case, $L$ and $N$ exhibit
diffusive behavior $\gamma_L=\gamma_N=\frac{1}{2}$,
but we observed  $\gamma_\rho=\frac{1}{4}$.
For the backward case, the exponents again
depend on the hopping parameter $p$.

The results presented here show surprisingly an
update-dependent critical behavior of the EQP. The critical
behavior of the EQP is nonuniversal in the sense that it depends on
the update rule and, for the backward update, the 
hopping parameter $p$.
Although there are many studies on the TASEP and related models 
with fixed system length, as far as we know, 
such update-dependent property has not been observed.
The  strong sensitivity to the details of the dynamics is rather 
unusual and requires further investigation.
We expect that stochastic particle systems with 
varying system size will be found to exhibit 
many other interesting phenomena.

\acknowledgments
The authors thank Kirone Mallick for useful discussions.
C Arita acknowledges support from the JSPS fellowship program for research abroad.

\end{document}